\documentclass[11pt,twocolumn]{extarticle}

\usepackage[letterpaper,top=0.52in,bottom=0.52in,left=0.62in,right=0.62in,columnsep=0.24in]{geometry}
\usepackage[T1]{fontenc}
\usepackage[utf8]{inputenc}
\usepackage{newtxtext,newtxmath}
\usepackage{microtype}
\usepackage{booktabs}
\usepackage{amsmath}
\usepackage{enumitem}
\usepackage{titlesec}
\usepackage{caption}
\usepackage[hidelinks]{hyperref}
\usepackage{balance}

\hypersetup{
  pdftitle={Measuring User Success in Randomized Search Experiments: Interpreting Good Abandonment in Generative AI Search},
  pdfauthor={Minha Hwang, Shavak Agrawal, David Chen, Carto Wong, Xin Du, Tiannan Wang, Georg Buscher, Zheng Chen}
}
\setlist{nosep,leftmargin=1.4em}
\titleformat{\section}{\normalfont\bfseries\fontsize{11.5}{12.5}\selectfont}{\thesection.}{0.35em}{}
\titlespacing*{\section}{0pt}{5pt plus 1pt minus 1pt}{2pt}

\title{\vspace{-2.3em}\bfseries\fontsize{15}{16.5}\selectfont
Measuring User Success in Randomized Search Experiments:\\
Interpreting Good Abandonment in Generative AI Search}
\author{\fontsize{9}{10}\selectfont
Minha Hwang\textsuperscript{1} \quad
Shavak Agrawal\textsuperscript{1} \quad
David Chen\textsuperscript{1} \quad
Carto Wong\textsuperscript{1}\\
\fontsize{9}{10}\selectfont
Xin Du\textsuperscript{1} \quad
Tiannan Wang\textsuperscript{1} \quad
Georg Buscher\textsuperscript{1} \quad
Zheng Chen\textsuperscript{2}\\[-0.15em]
\fontsize{9}{10}\selectfont
\textsuperscript{1}Microsoft \qquad \textsuperscript{2}Simplex AI}
\date{}

\begin{document}
\maketitle
\vspace{-2.2em}

\begin{abstract}
\noindent
Randomized search experiments increasingly evaluate experiences in which large language models answer information needs directly on the results page. Click status alone cannot identify success: a click may lead to an unsatisfactory result, while no click may indicate successful answer consumption, or ``good abandonment.'' Prior research has shown that good abandonment is difficult to detect because dwell, scrolling, and reformulation are meaningful only in context. We present Session Success Rate (SSR), a human-grounded, interpretable outcome metric designed for online controlled experiments. Trained judges label complete session experiences using privacy-protected interaction evidence and video recordings; these offline labels train models scored online from query events, viewport behavior, result-page interactions, and semantic reformulations. Out-of-sample evaluation on 6,904 labeled queries yields 0.75 balanced accuracy overall and 0.68 within abandoned queries; session-level balanced accuracy is 0.68. The sample shows why interaction status is not an adequate outcome: 53.2\% of abandonments were successful and 6.5\% of clicked queries were unsuccessful. In a validation portfolio of seven ranking and five non-ranking experiments, SSR agreed with Sessions per User in seven cases, was inconclusive in four, and disagreed in one under a common \(p<0.05\) rule. Separately, across approximately 4,000 production experiments, SSR produced about 2.5 times the binary sensitivity of Sessions per User; in the studied workflow, this supported lower traffic allocations and greater experiment throughput under a fixed traffic budget.
\end{abstract}

\section{Experimentation Problem}

Digital experiments commonly use clicks, dwell time, lexical reformulation, or longer-run activity as proxies for search success \cite{kohavi2020,gupta2019}. Generative answers make their limitations more consequential. Click behavior does not map uniformly to retrieval quality \cite{radlinski2008}; a click can precede dissatisfaction, while an on-page answer can satisfy a user without a click. Research on good abandonment has therefore emphasized that abandonment cannot be interpreted without contextual behavioral evidence \cite{li2009,williams2016}. Long dwell, for example, may indicate answer consumption or difficulty depending on scrolling and subsequent reformulation \cite{hassan2014}.

This creates an experimental-measurement problem rather than merely a classification problem: can a human-grounded behavioral outcome improve the interpretation of randomized search experiments when both clicks and abandonments contain successful and unsuccessful experiences? SSR is designed around three principles established during development: \textbf{grounding} in human judgments, \textbf{interpretability} for experiment diagnosis, and \textbf{robustness} against feature-specific signals and treatment-dependent telemetry. The contribution is an operational framework for validating, deploying, and diagnosing a human-grounded experimental outcome when clicks and abandonments each contain both successful and unsuccessful experiences.

\section{Human Ground Truth and Evaluation Design}

Development follows five stages: stratified session sampling, human labeling, behavioral-signal curation, query- and session-level modeling, and offline plus experiment-level evaluation. A session is delimited by 30 minutes of inactivity, following a standard industry convention. Sessions are sampled across device type, session length, and broad query-frequency segments. Multi-query and mobile sessions are oversampled because they are harder to interpret and underrepresented in unadjusted traffic; sessions are selected randomly within strata. Population prevalence requires inverse sampling weights.

Trained judges review privacy-protected query sequences, result interactions, page observations, dwell patterns, and video recordings of the complete search experience. Video is used only to construct offline ground truth, not as an online scoring input. At least two query- and session-level judgments are collected per item.

\begin{table}[t]
\centering
\caption{Human-labeled outcome sample.}
\label{tab:sample-v2}
\scriptsize
\setlength{\tabcolsep}{2.2pt}
\begin{tabular}{lrrr}
\toprule
Unit & Successful & Unsuccessful & Total \\
\midrule
Clicked queries & 4,684 (93.5\%) & 325 (6.5\%) & 5,009 \\
Abandoned queries & 1,008 (53.2\%) & 887 (46.8\%) & 1,895 \\
All queries & 5,692 (82.4\%) & 1,212 (17.6\%) & 6,904 \\
Sessions & 2,742 (85.3\%) & 474 (14.7\%) & 3,216 \\
\bottomrule
\end{tabular}
\vspace{1pt}
\parbox{\columnwidth}{\scriptsize\emph{Note:} The sample oversamples difficult sessions and is not a population prevalence estimate.}
\end{table}

Table~\ref{tab:sample-v2} demonstrates the construct problem. A click-status rule necessarily labels every click alike and every abandonment alike; it cannot distinguish the 325 unsuccessful clicked queries from satisfactory clicks or the 1,008 successful abandonments from failed abandonments. It is therefore not an appropriate baseline for the intended outcome, which requires discrimination \emph{within} both interaction classes. We instead report class-balanced performance against human success labels and compare the abandonment branch with the prior abandonment rule.

All reported model performance is out of sample. All queries from a session remain in the same partition, and model fitting and threshold selection exclude the evaluated partition. Because success is substantially more common than failure, balanced accuracy is primary; raw accuracy is included to show the effect of class imbalance (Table~\ref{tab:performance-v2}).

\begin{table}[t]
\centering
\caption{Out-of-sample performance against human success judgments.}
\label{tab:performance-v2}
\scriptsize
\setlength{\tabcolsep}{1.5pt}
\begin{tabular}{@{}p{0.52\columnwidth}rrr@{}}
\toprule
Model / baseline & \(N\) & Acc. & Bal. acc. \\
\midrule
Majority-class query baseline & 6,904 & 0.824 & 0.500 \\
Query success model & 6,904 & 0.810 & 0.750 \\
Prior abandonment rule & 1,895 & 0.450 & 0.450 \\
SSR abandonment branch & 1,895 & 0.680 & 0.680 \\
Majority-class session baseline & 3,216 & 0.853 & 0.500 \\
Session success model & 3,216 & 0.890 & 0.680 \\
\bottomrule
\end{tabular}
\end{table}

\section{Interpretable Measurement Model}

SSR begins at the query level so experiment movements can be decomposed. Inputs combine: (1) \textbf{event signals}, including query order, clicks, downstream engagement, and timing; (2) \textbf{page-interaction signals}, including result-page dwell, relative visibility, and scrolling; and (3) \textbf{semantic journey signals}, especially struggle-indicating reformulations.

The decision structure separates clicked and abandoned queries. For clicks, longer downstream engagement, limited result-page scanning, and no struggling reformulation support success; brief downstream engagement followed by continued struggle supports failure. For abandonment, meaningful result-page engagement, limited deep scanning, no immediate downward scroll, and no struggling reformulation support success. No signal is sufficient alone. Joint behavioral conditions prevent, for example, long dwell from being interpreted identically when it is accompanied by extensive scanning and reformulation.

We compare whether any query succeeds, whether the final query succeeds, and average query success. Average success performs best among these transparent rules. For session \(s\), with \(n_s\) query predictions \(\hat y_{si}\),
\begin{equation}
\mathrm{SSR}_s=\frac{1}{n_s}\sum_{i=1}^{n_s}\hat y_{si}.
\end{equation}
Additional controls yield only marginal predictive gains while reducing auditability and robustness, so the simpler aggregation is retained.

\section{Randomized-Experiment Evidence}

SSR is evaluated against Sessions per User across 12 online controlled experiments: seven ranking or result-presentation interventions and five non-ranking interventions. The NSA validation set is restricted to experiments in which Sessions per User moves significantly and is therefore treated as the engagement ground truth for directional comparison. We use a common two-sided \(p<0.05\) rule and a three-state Net Score of Agreement (NSA) classification: \emph{agree} when significant SSR movement has the same direction, \emph{inconclusive} when SSR does not move significantly, and \emph{disagree} when significant SSR movement has the opposite direction. Seven experiments agree, four are inconclusive, and one disagrees (Table~\ref{tab:experiments-v2}). Concordance is strongest for ranking interventions and weaker for broader interface or monetization changes. These comparisons assess directional consistency with the engagement reference; they do not by themselves prove that either metric identifies latent satisfaction.

\begin{table}[t]
\centering
\caption{Directional comparison across randomized experiments.}
\label{tab:experiments-v2}
\scriptsize
\setlength{\tabcolsep}{4pt}
\begin{tabular}{lrrr}
\toprule
Intervention family & Agree & Inconclusive & Disagree \\
\midrule
Ranking / presentation & 6 & 1 & 0 \\
Other interventions & 1 & 3 & 1 \\
\midrule
All experiments & 7 & 4 & 1 \\
\bottomrule
\end{tabular}
\end{table}

\textbf{Sanitized experiment examples.}
In one ranking-model experiment, click indicators improved, but SSR-predicted good abandonment declined. A click-centric metric suggested a clear gain, whereas SSR's clicked-versus-abandoned decomposition produced no significant aggregate improvement and reframed the result as a tradeoff. In a second ranking-weight experiment, a legacy click-centric indicator regressed even though SSR-predicted good abandonment improved and longer-run engagement did not conflict. SSR captured the abandonment gain and reframed the apparent failure as a mixed result. In both cases, the resulting decision was to investigate and iterate rather than immediately ship or reject the treatment. The examples show how SSR guards against overconfidence from click gains and premature rejection when success shifts toward on-page consumption.

Individual experiment deep dives are conducted jointly with subject-matter experts from the owning product area, such as ranking teams. Reviewers examine the treatment mechanism, expected user impact, clicked and abandoned success components, reformulation behavior, platform differences, and established engagement outcomes. These structured reviews assess whether SSR adds decision-relevant information beyond legacy metrics and identify cases requiring additional validation rather than mechanically privileging the new metric.

Operational validity is checked before interpreting an effect. SSR is not used when treatment changes the availability of required page-observation or semantic inputs. Metric-level treatment-control coverage and sample-ratio checks detect such cases; fallback reformulation measures or established outcomes are used instead. This safeguard is important because interface treatments can mechanically change viewport, scrolling, dwell, or representation availability.

In a separate production portfolio of approximately 4,000 experiments, SSR provides about 2.5 times the \emph{binary sensitivity} of Sessions per User, where binary sensitivity is the number of experiments with statistically significant metric movement divided by the total number of experiments under the same \(p<0.05\) rule. Both metrics are compared on the same eligible experiments, observation windows, and inference procedures; experiments with treatment-control imbalance in required SSR telemetry are excluded. The ratio is rounded and reported relatively because raw production counts are confidential. In the studied workflow, the higher detection frequency supported lower traffic allocations per experiment, greater throughput under the same traffic budget, shorter time-to-decision, and faster shipping decisions. The significant-result-rate ratio is not an equal-power sample-size conversion, and these workflow benefits do not isolate the causal effect of adopting SSR. SSR is deployed alongside established user-success outcomes rather than as a stand-alone source of truth.

\section{Diagnostic Value and Limitations}

Query Success Rate (QSR), the query-level companion to SSR, enables experimenters to identify which normalized query segments show statistically significant treatment-control movement and which segments contribute most to the aggregate effect. Segment-level testing applies multiple-comparison correction. This supports targeted debugging, representative-query review, and analysis of heterogeneous user impact while preserving the randomized comparison \cite{athey2017}. The segment decomposition is diagnostic rather than a separate causal effect estimate for every category.

QSR success and failure are partitioned into mutually exclusive clicked-query and abandonment modes. Its decision tree traces the flow of queries through nodes and branches defined by engagement, page visibility, reformulation, scrolling, and dwell, connecting an aggregate movement to its behavioral mechanism. Per-user contribution views further distinguish outcome changes from query-volume or denominator changes and connect online movers with query-level offline evaluation.

Limitations remain. Human judges infer success from observable session evidence rather than directly surveying users. Validity depends on stable relationships between behavior and success, so new interfaces require coverage checks, monitoring, and periodic human-label review. Evidence is strongest for ranking and result-presentation experiments; broader intervention classes need further validation. Finally, the data represent evolving answer-oriented search rather than an isolated generative-only population.

\section{Conclusion}

Good abandonment has remained difficult to measure because neither a click nor an abandonment has a uniform success meaning. SSR addresses this experimental blind spot by learning from complete, human-reviewed sessions while retaining interpretable online signals and explicit validity gates. Out-of-sample results show discrimination within abandonment as well as overall query and session performance, while 12 randomized experiments provide initial evidence of directional alignment. In a separate production workflow, about 2.5 times greater binary sensitivity supported lower per-experiment traffic allocation and greater throughput under a fixed traffic budget. The framework offers experimenters a practical bridge between offline human judgment, online causal comparisons, and diagnosis of AI-powered search changes.

\balance


\begin{thebibliography}{7}
\setlength{\itemsep}{0pt}
\setlength{\parskip}{0pt}
\footnotesize

\bibitem{athey2017}
S. Athey and G. W. Imbens. 2017.
The econometrics of randomized experiments.
In \emph{Handbook of Economic Field Experiments}.

\bibitem{gupta2019}
S. Gupta, R. Kohavi, D. Tang, Y. Xu, R. Andersen, E. Bakshy, N. Cardin, N. Chandran, N. Chen, D. Coey, et al. 2019.
Top challenges from the first practical online controlled experiments summit.
\emph{SIGKDD Explorations}.

\bibitem{hassan2014}
A. Hassan, R. W. White, S. T. Dumais, and Y.-M. Wang. 2014.
Struggling or exploring? Disambiguating long search sessions.
In \emph{WSDM}, 53--62.

\bibitem{kohavi2020}
R. Kohavi, D. Tang, and Y. Xu. 2020.
\emph{Trustworthy Online Controlled Experiments: A Practical Guide to A/B Testing}.
Cambridge University Press.

\bibitem{li2009}
J. Li, S. B. Huffman, and A. Tokuda. 2009.
Good abandonment in mobile and PC internet search.
In \emph{SIGIR}, 43--50.

\bibitem{radlinski2008}
F. Radlinski, M. Kurup, and T. Joachims. 2008.
How does clickthrough data reflect retrieval quality?
In \emph{CIKM}.

\bibitem{williams2016}
K. Williams, J. Kiseleva, A. C. Crook, I. Zitouni, A. Hassan Awadallah, and M. Khabsa. 2016.
Detecting good abandonment in mobile search.
In \emph{WWW}.

\end{thebibliography}
\end{document}